\documentclass[journal]{IEEEtran}

\usepackage{amsmath,amssymb,amsthm}
\usepackage{xifthen}
\usepackage{mathtools}
\usepackage{enumerate}
\usepackage{microtype}
\usepackage{xspace}
\usepackage{bm}
\usepackage[T1]{fontenc}
\usepackage{fancyhdr}
\usepackage{lastpage}
\usepackage{bbm}

\newcommand{\mbs}[1]{\bm{#1}}
\newcommand{\vect}[1]{{\lowercase{\mbs{#1}}}}
\newcommand{\mat}[1]{{\uppercase{\mbs{#1}}}}

\newcommand{\T}{{\scriptscriptstyle\mathsf{T}}}
\renewcommand{\H}{{\scriptscriptstyle\mathsf{H}}}

\renewcommand{\Re}[1][]{\ifthenelse{\isempty{#1}}{\operatorname{Re}}{\operatorname{Re}\left(#1\right)}}
\renewcommand{\Im}[1][]{\ifthenelse{\isempty{#1}}{\operatorname{Im}}{\operatorname{Im}\left(#1\right)}}

\newcommand{\gv}{\vect{g}}
\newcommand{\hv}{\vect{h}}

\newcommand{\sv}{\vect{s}}

\newcommand{\wv}{\vect{w}}

\newcommand{\yv}{\vect{y}}
\newcommand{\zv}{\vect{z}}

\newcommand{\alphav}{\vect{\alpha}}

\newcommand{\upsilonv}{\vect{\upsilonv}}

\newcommand{\Phim}{\mat{\Phi}}
\newcommand{\Psim}{\mat{\Psi}}

\newcommand{\Upsilonm}{\mat{\Upsilon}}

\newcommand{\Gm}{\mat{g}}

\newcommand{\Qm}{\mat{q}}

\newcommand{\Ac}{{\mathcal A}}

\newcommand{\Cc}{{\mathcal C}}

\newcommand{\Nc}{{\mathcal N}}

\newcommand{\Pc}{{\mathcal P}}

\newcommand{\CC}{\mathbb{C}}

\newcommand{\Id}{\mat{\mathrm{I}}}

\newcommand{\CN}[1][]{\ifthenelse{\isempty{#1}}{\mathcal{N}_{\mathbb{C}}}{\mathcal{N}_{\mathbb{C}}\left(#1\right)}}

\renewcommand{\P}[1][]{\ifthenelse{\isempty{#1}}{\mathbb{P}}{\mathbb{P}\left(#1\right)}}
\newcommand{\E}[1][]{\ifthenelse{\isempty{#1}}{\mathbb{E}}{\mathbb{E}\left[#1\right]}}
\newcommand{\I}[1][]{\ifthenelse{\isempty{#1}}{\mathbb{I}}{\mathbb{I}\left\{#1\right\}}}
\renewcommand{\det}[1][]{\ifthenelse{\isempty{#1}}{\mathrm{det}}{\text{det}\left(#1\right)}}
\newcommand{\trace}[1][]{\ifthenelse{\isempty{#1}}{\mathrm{tr}}{\text{tr}\left(#1\right)}}
\newcommand{\rank}[1][]{\ifthenelse{\isempty{#1}}{\mathrm{rank}}{\text{rank}\left(#1\right)}}
\newcommand{\diag}[1][]{\ifthenelse{\isempty{#1}}{\mathrm{diag}}{\text{diag}\left(#1\right)}}
\newcommand{\Cov}[1][]{\ifthenelse{\isempty{#1}}{\mathsf{Cov}}{\mathsf{Cov}\left(#1\right)}}

\newcommand{\defeq}{\triangleq}

\newcounter{enumi_saved}
\usepackage{answers}
\Newassociation{solution}{Solution}{solutionfile}

\AtBeginDocument{\Opensolutionfile{solutionfile}[\jobname]}
\AtEndDocument{\Closesolutionfile{solutionfile}\clearpage
}

\IfFileExists{MinionPro.sty}{
}{
}

\usepackage{algorithm,algorithmic}

\usepackage{cite}
\usepackage{subfigure,graphics}
\usepackage{multirow}
\usepackage{array}
\usepackage{amsmath,bbm}
\usepackage{color}
\usepackage{enumitem}
\usepackage{stfloats}
\usepackage[flushleft]{threeparttable}
\usepackage{upgreek}
\usepackage{pgfplots}
\pgfplotsset{minor grid style={dotted,gray!25}}
\pgfplotsset{major grid style={dashed,gray!25}}

\usepackage{grffile}
\pgfplotsset{compat=newest}
\usetikzlibrary{plotmarks}
\usetikzlibrary{arrows.meta}
\usepgfplotslibrary{patchplots}
\newcommand{\ind}[1]{{\mathbbm{1}{\{#1\}}}}

\renewcommand{\defeq}{\triangleq}

\DeclareMathOperator*{\minimize}{minimize}
\DeclareMathOperator*{\maximize}{maximize}

\newcommand{\hEN}[1][]{\ifthenelse{\isempty{#1}}{\hv_{{\rm EN}}}{\hv_{{\rm EN},#1}}}
\newcommand{\hR}[1][]{\ifthenelse{\isempty{#1}}{\hv_{{\rm H}}}{\hv_{{\rm H},#1}}}
\newcommand{\hE}[1][]{\ifthenelse{\isempty{#1}}{\hv_{{\rm EVE}}}{\hv_{{\rm EVE},#1}}}
\newcommand{\hathE}[1][]{\ifthenelse{\isempty{#1}}{\hat{\hv}_{{\rm EVE}}}{\hat{\hv}_{{\rm EVE},#1}}}
\newcommand{\tildehE}[1][]{\ifthenelse{\isempty{#1}}{\tilde{\hv}_{{\rm EVE}}}{\tilde{\hv}_{{\rm EVE},#1}}}
\newcommand{\REN}[1][]{\ifthenelse{\isempty{#1}}{R}{R_{#1}}}
\newcommand{\RE}[1][]{\ifthenelse{\isempty{#1}}{R_{{\rm EVE}}}{R_{{\rm EVE},#1}}}
\newcommand{\barRE}[1][]{\ifthenelse{\isempty{#1}}{\bar{R}_{{\rm EVE}}}{\bar{R}_{{\rm EVE},#1}}}
\newcommand{\RS}[1][]{\ifthenelse{\isempty{#1}}{R_{{\rm s}}}{R_{{\rm s},#1}}}

\usepackage{tikz}
\usetikzlibrary{calc,fadings,decorations.pathreplacing}

\tikzset{%
	>=latex, % option for nice arrows
	inner sep=0pt,%
	outer sep=2pt,%
	mark coordinate/.style={inner sep=0pt,outer sep=0pt,minimum size=3pt,
		fill=black,circle}%
}

\title{Low-Latency and Secure Computation Offloading Assisted by Hybrid Relay-Reflecting Intelligent Surface} 
  
\author{ \IEEEauthorblockN{Khac-Hoang Ngo\IEEEauthorrefmark{1}, Nhan Thanh Nguyen\IEEEauthorrefmark{2}, Thinh Quang Dinh\IEEEauthorrefmark{3}, Trong-Minh Hoang\IEEEauthorrefmark{4}, and Markku Juntti\IEEEauthorrefmark{2}
}
\IEEEauthorblockA{\IEEEauthorrefmark{1}Department of Electrical Engineering, Chalmers University of Technology, 41296 Gothenburg, Sweden \\
\IEEEauthorrefmark{2}Centre for Wireless Communications,
University of Oulu, P.O.Box 4500, FI-90014, Finland \\
\IEEEauthorrefmark{3}Center of Excellence for Wearables Research and Development, Fossil Group, Inc., Vietnam\\
\IEEEauthorrefmark{4}Posts and Telecommunications Institute of Technology, Hanoi, Viet Nam\\
{\small Emails: ngok@chalmers.se, \{nhan.nguyen, markku.juntti\}@oulu.fi, thinhdinh@ieee.org, hoangtrongminh@ptit.edu.vn}}
\vspace{-.8cm}
}

\begin{document}

\maketitle
\thispagestyle{plain}

\begin{abstract} 
    Recently, the hybrid relay-reflecting intelligent surface (HRRIS) has been introduced as a spectral- and energy-efficient architecture to assist wireless communication systems. In the HRRIS, a single or few active relay elements are deployed along with a large number of passive reflecting elements, allowing it to not only reflect but also amplify the incident signals. In this work, we investigate the potential of the HRRIS in aiding the computation offloading in a single-user mobile edge computing system. The objective is to minimize the offloading latency while ensuring the secrecy of user data against a malicious eavesdropper. We develop efficient solutions to this latency minimization problem based on alternating optimization. Through numerical results, we show that the deployment of the HRRIS can result in a considerable reduction in latency. Furthermore, the latency reduction gain offered by the HRRIS is much more significant than that of the conventional reconfigurable intelligent surface (RIS).
\end{abstract}
\vspace{-.3cm}
% \begin{IEEEkeywords}
% 	
% \end{IEEEkeywords}

%-----------------------------------------------
\section{Introduction} \label{sec:intro}
Edge computing is a novel paradigm to develop communication and computation infrastructures for the Internet of Things. It overcomes the high latency and low bandwidth drawbacks of centralized cloud computing by extending the cloud's capacities to near-user network edges \cite{LA20193}. In a mobile edge computing~(MEC) system, the mobile users offload their entire or partial computation tasks to proximate MEC servers via wireless links. Thus, computation offloading %involves the optimization of not only the offloading volumes and computational resource allocation but also the user-server communications. This communications 
suffers from fading and attenuation in the wireless medium~\cite{Jiang2019toward}. This may entail a higher latency than the local execution, and hinder the advantage of MEC, especially for users located far from edge nodes~\cite{Bai2020latency}. Therefore, efficient methods to improve the user-server communications are needed.

A new technology called reconfigurable intelligent surface (RIS) has emerged as a promising solution to enhance the wireless communication capacity by reflecting radio waves in preferred directions~\cite{di2020smart, he2021channel, wu2020intelligent, nguyen2021machine}. RIS is comprised of a controller and a large number of passive reflecting elements which enable passive beamforming without requiring any radio frequency (RF) chains~\cite{liu2021reconfigurable, wu2020intelligent}. However, the main limitation of RIS is that the passive reflection limits the beamforming gains. Recently, a novel concept of hybrid relay-reflecting intelligent surface~(HRRIS) has been proposed to overcome this limitation~\cite{nguyen2021hybrid_mag, nguyen2021spectral, nguyen2021hybrid}. In the HRRIS, a few elements are equipped with power amplifiers to serve as active relays. It was shown that HRRIS, even with a single active element, significantly improves both the system spectral and energy efficiency (SE/EE) with respect to RIS~\cite{nguyen2021hybrid_mag, nguyen2021spectral, nguyen2021hybrid}. Due to its potential, HRRIS can be deployed to assist the computation offloading in MEC. RIS-assisted MEC system has been investigated recently in, e.g.,~\cite{Bai2020latency,Hu2021reconfigurable}. However, to our best knowledge, the applications of HRRIS for computation offloading have not been considered in the literature.

Besides latency minimization, another challenge in MEC computation offloading is to guarantee the secrecy of user data under the presence of malicious eavesdroppers in the system. Various techniques have been proposed to maximize the secrecy communication rate such as jamming with artificial noise and beamforming schemes \cite{khisti2010secure, liu2014secrecy, mukherjee2014principles}. However, when the legitimate communication channel is weaker than the eavesdropping channel, the achievable secrecy rate is significantly limited, even with the aforementioned techniques \cite{cui2019secure}. RIS has been shown to be an efficient solution to tackle this challenge \cite{cui2019secure, chen2019intelligent, hong2020artificial, yu2020robust, shen2019secrecy}. However, with the powerful reflecting/relaying capabilities, HRRIS is expected to provide significant improvement in the secrecy rate compared with the conventional passive RIS.

In this paper, we aim at minimizing the latency in computation offloading while guaranteeing secrecy in a single-user HRRIS-aided MEC system. We assume that the user has a local computing resource and offloads a fraction of the tasks to an edge node~(EN). A multi-antenna eavesdropper receives the user's signal and attempts to decode the data. To prevent that, the EN informs the user to transmit at a secrecy rate lower than the maximal achievable rate. For the latency minimization, we propose a joint design of the receive combining vector, the HRRIS's reflecting/relaying coefficients (communication parameters), and the computation offloading volume (computation parameter). To this end, we develop an alternating optimization approach that efficiently solves the challenging latency minimization problem. In particular, we consider both the fixed HRRIS and the dynamic HRRIS schemes. In the former, the positions and number of the active elements are unchanged; in contrast, those in the latter can be dynamically changed and optimized according to the channel condition. We numerically evaluate the latency of these schemes and compare it with that of the conventional RIS. Our simulation results show that the dynamic HRRIS consistently outperforms the fixed HRRIS, and both provide significant latency reduction with respect to RIS. % with random or optimized phase, and investigate the variation of the latency with respect to various system parameters, namely, the location of the user, the number of elements, the local/edge computing capability, the number of eavesdropper's antennas, and the power budget of the HRRIS. 

\section{System Model} \label{sec:model}

\begin{figure}[t!]
	\vspace{-1cm}
	\centering
	\includegraphics[width=.45\textwidth]{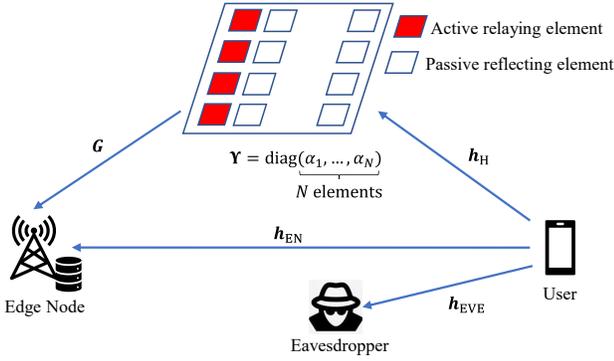}
	\vspace{-1cm}
	\caption{The HRRIS-assisted communication and computation offloading system in the presence of an eavesdropper.}
	\vspace{-0.3cm}
	\label{fig:model}
\end{figure}

We consider a single-user system in the presence of a malicious eavesdropper, as illustrated in Fig.~\ref{fig:model}. The user has computation tasks to partially offload to an EN. An HRRIS is deployed in the system to assist the offloading so that the computation result is returned to the user with low latency, while the user data is protected against the eavesdropper. The model is detailed in the following.
 
\subsection{Communication Model}

We assume that the user, the EN, and the eavesdropper are equipped with a single antenna, $M$ antennas, and $E$ antennas, respectively.  %We further assume that the communication between the EN and users is assisted by a hybrid relay-reflecting intelligent surface~(HRRIS)~\cite{nguyen2021hybrid}. 
The HRRIS has $N$ elements, including $A$ active relaying elements and $N\!-\!A$ passive reflecting elements. We note that a passive reflecting coefficient has the optimal amplitude of unity to maximize the received signal power \cite{wu2018intelligent}; therefore, a passive element only adjusts the incident signal's phase. In contrast, an active one can not only tune the phase but also amplify the signal power. We denote the index set of the positions of the $A$ active elements by $\Ac \subset [N] \defeq \{1,2,\dots,N\}$. Let $\alpha_n = |\alpha_n| e^{\jmath \theta_n}$ denote the relay/reflection coefficient of the $n$-th element, where $|\alpha_n|$ and $\theta_n \in [\pi,2\pi)$ represent the amplitude and phase shift, respectively. Note that $|\alpha_n| = 1$, $\forall n \notin \Ac$. Let $\alphav \defeq [\alpha_1 \dots \alpha_N]^\T$. We denote $\Upsilonm \!\defeq\! \diag[\alpha_1,\dots,\alpha_N]$ and decompose $\Upsilonm$ as $\Upsilonm = \Phim + \Psim$ where $\Phim = \diag[\phi_1,\dots,\phi_N]$ and $\Psim = \diag[\psi_1,\dots,\psi_N]$ with $\phi_n = \alpha_n \ind{n\notin \Ac}$ and $\psi_n = \alpha_n \ind{n\in \Ac}$, where $\ind{\cdot}$ is the indicator function. That is, $\Phim$ and $\Psim$ contain the passive and
active coefficients, respectively.

%which is given by
%\begin{align}
%	\alpha_n = \begin{cases}
%		|\alpha_n| e^{\jmath \theta_n}, &\text{if $n\in \Ac$}, \\
%		e^{\jmath \theta_n}, &\text{otherwise},
%	\end{cases}
%\end{align}
%where $\theta_n \in [\pi,2\pi)$ represent the phase shift.

Let $\hEN \!\in\! \CC^{M}$, $\hR \!\in\! \CC^{N}$, and $\hE \!\in\! \CC^{E}$ denote the channel vectors between the user and the EN, the HRRIS, and the eavesdropper, respectively. Let $\Gm \in \CC^{M\times N}$ denote the channel matrix from the EN to the HRRIS.  We assume that these channel coefficients are quasi-static, i.e., remain unchanged during the whole computation offloading circle. Furthermore, we assume that the eavesdropper knows $\hE$; the EN knows perfectly $\hEN$, $\hR$, and $\Gm$ but only knows an estimate $\hathE$ of $\hE$. Following the deterministic uncertainty model \cite{Khandaker2018,He2020}, we assume that 
%\begin{align} \label{eq:CSI_uncertainty}
	$\frac{\|\hE - \hathE\|}{ \|\hathE\|} \le \epsilon$,
%\end{align}
where $\epsilon$ is an upper bound of the relative
estimation error.

Denote the transmit power and transmitted signal of the user by $P$ and $s$, respectively. The incident signal at the HRRIS is $\sqrt{P} \hR s + \zv_{\rm H}$, where $\zv_{\rm H} \sim \Cc\Nc(0,\sigma_{\rm H}^2 \Id_N)$ is the noise vector. The transmit power of the HRRIS's active elements is given as
%\begin{align}
	$P_{\rm a}(\alphav) \defeq \trace[{\Psim(P \hR \hR^\H + \sigma^2_{\rm H} \Id_N) \Psim^\H}]$.
%\end{align}
The received signal at the EN is given by
\begin{align}
	\yv &= \sqrt{P} (\hEN + \Gm \Upsilonm \hR) s_k + \Gm \Psim \zv_{\rm H} + \zv_{\rm EN} \\
	&= \sqrt{P} \hv \sv + \zv,
\end{align}
where $\hv \defeq \hEN + \Gm \Upsilonm \hR$ denotes the effective user-EN channel, $\zv_{\rm EN} \!\sim\! \Cc\Nc(0,\sigma_{\rm EN}^2 \Id_M)$ is the additive noise at the EN, and $\zv \defeq \Gm \Psim \zv_{\rm H} + \zv_{\rm EN}$ is the total effective noise. For notational simplicity, we assume that $\sigma^2_{\rm H} = \sigma^2_{\rm EN} = \sigma^2$, and thus $\zv \sim \Cc\Nc\left(0,\sigma^2 \Qm \right)$ where $\Qm \defeq \Id_M + \Gm\Psim\Psim^\H\Gm^\H$.

We assume that the EN employs a linear combining vector $\wv \!\in\! \CC^{M}$, and thus, the signal is recovered at the EN as $\hat{s} = \wv^\H \yv$.
%\begin{align}
%	\hat{s} = \wv^\H \yv = \Wm^\H (\sqrt{P} \Hm \sv + \zv).
%\end{align}
%Specifically, the recovered signal for user $k$ is
%\begin{align}
%	$\hat{s} = \wv^\H (=\sqrt{P} \sum_{j=1}^{K} \hv_j s_j + \zv\right]$.
%\end{align}
Then, the effective signal-to-interference-plus-noise ratio~(SINR) is given by 
$
	\gamma(\wv,\alphav) = \frac{P |\wv^\H\hv|^2}{\sigma^2 \wv^\H\Qm \wv}.
$
Therefore, the maximal achievable rate is given by
\begin{align}
	\REN(\wv,\alphav) = W \log_2[1+\gamma(\wv_k,\alphav)], 
\end{align}
where $W$ is the uplink bandwidth.

The eavesdropper also receives the user's signal and attempts to decode the data.\footnote{We assume that the eavesdropper does not receive the signal from the HRRIS.} The rate that the eavesdropper can achieve from user $k$'s signal, i.e., the leakage rate, is
$
	\RE = W \log_2\left(1+\frac{P\|\hE\|^2}{\sigma^2_{\rm EVE}}\right),
$
where $\sigma^2_{\rm EVE}$ is the noise variance at the eavesdropper.
The secrecy rate for user $k$ is then given by $\REN(\wv,\alphav) - \RE$. Since the EN only knows $\hathE$, it computes an upper bound $\barRE$ on the leakage rate:
\begin{align}
	\barRE \defeq \log_2\left(1+\frac{P(1+\epsilon)^2\|\hathE\|^2}{\sigma^2_{\rm EVE}}\right).
\end{align}
%\begin{align}
%	\RE[k] &\le \log_2\left(1+\frac{P(1+\epsilon)^2\|\hathE[k]\|^2}{P(1-\epsilon)^2\sum_{j=1,j\ne k}^{K} \|\hathE[j]\|^2 + \sigma^2_{\rm E}}\right) \label{eq:tmp196}\\
%	&\eqdef \barRE[k].
%\end{align} 
%The inequality in \eqref{eq:tmp196} follows from \eqref{eq:CSI_uncertainty} and the triangle inequality. 
To guarantee secrecy, the EN informs the user to transmit at a lower bound on the secrecy rate which is given by
\begin{align}
	\RS(\wv,\alphav) = (\REN(\wv,\alphav) - \barRE)^+.
\end{align}

\subsection{Computing Model}
The user has some computational tasks and might offload a certain fraction or all of their tasks to the EN. Thus, $s$ is the offloading signal. %The computing model is detailed as follows. 
Let $L$, $\ell$, and $\nu$ denote the total number of bits to be processed, the number of bits offloaded to the edge server, and the number of CPU cycles required to process one bit, respectively.
\begin{itemize}[leftmargin=*]
	\item \textit{Local computing:} Denote the computational capability of the user by $f^{l}$ CPU cycles/second. Then the time required for the local computation at user~$k$ is given by 
	$
		D^{l} = \frac{(L - \ell)\nu}{f^l}.
	$
	
	\item \textit{Edge computing:} We denote the computational capability of the MEC as~$f^e$  CPU cycles/second. Since the computation results are typically of small size, the delay due to feedback of these results to the user is assumed to be negligible. Therefore, the total latency imposed by the computation offloading and the edge computing is given by 
	$
		D^{e} = \frac{\ell}{\RS(\wv,\alphav)} + \frac{\ell \nu}{f^e}. 
	$
\end{itemize}
The overall latency is imposed by the maximum latency between the local and edge computing, that is, $$D(\wv,\alphav,\ell) = \max\{D^{l}, D^{e}\}.$$
%\begin{align}
%	D(\wv,\alphav,\ell,f^e) &= \max\{D^{l}, D^{e}\} \\
%	&= \max\left\{\frac{(L_k - \ell_k)c_k}{f_k^l}, \frac{\ell_k}{R_k(\wv_k,\alphav)} + \frac{\ell_k c_k}{f_{k}^e}\right\}.
%\end{align}

\subsection{Problem Formulation}
We aim at minimizing the offloading latency by jointly optimizing the computation offloading volume $\ell$, the combining vector $\wv$, and the HRRIS's coefficients $\alphav$. In the fixed HRRIS architecture, the positions of the active elements, i.e., $\Ac$, are fixed and known. Thus, the latency minimization problem is
\begin{subequations}
\begin{align}
	(\Pc_{\rm fixed}) ~~	\minimize_{\wv, \alphav, \ell} &~ D(\wv,\alphav,\ell) \\
	\text{subject to} ~~%&\theta_n \in [0,2\pi), n \in N \\
	&|\alpha_n| = 1, n \notin \Ac, \label{eq:constraint_alpha}\\
	&P_{\rm a}(\alphav) \le P_{\rm a}^{\max}, \label{eq:constraint_Pa}\\
	&\ell \in \{0,1,\dots,L\}, \label{eq:constraint_ell}
%	&\sum_{k=1}^{K} f_{k}^e \le f_{\rm total}^e, \label{eq:constraint_ftot}\\
%	&f_k^e \ge 0, k \in [K], \label{eq:constraint_fe}
\end{align}
\end{subequations}
where $P_{\rm a}^{\max}$ is the power budget of the HRRIS. Considering the dynamic HRRIS scheme, $\Ac$ is unknown and thus, it is cast as a design parameter in the latency minimization with the dynamic HRRIS, i.e.,
\begin{subequations}
	\begin{align}
		(\Pc_{\rm dynamic}) ~~	\minimize_{\wv, \Ac, \alphav, \ell} &~ D(\wv,\alphav,\ell) \\ 
		\text{subject to} ~~&\eqref{eq:constraint_alpha}, \eqref{eq:constraint_Pa}, \eqref{eq:constraint_ell}, \notag \\
		&|\Ac| \le A, \Ac \subset [N]. \label{eq:constraint_A} 
		%	&\sum_{k=1}^{K} f_{k}^e \le f_{\rm total}^e, \label{eq:constraint_ftot}\\
		%	&f_k^e \ge 0, k \in [K], \label{eq:constraint_fe}
	\end{align}
\end{subequations}
In both problems $(\Pc_{\rm fixed})$ and $(\Pc_{\rm dynamic})$, $\wv$ and $\alphav$ are coupled. Furthermore, the objective function $D(\wv,\alphav,\ell)$ is segmented and nonconvex. Therefore, finding optimal solution is challenging. In the following, we develop efficient solutions to these optimization problems. 
 
%%%%%%%%%%%%%%%%%%%%%%%%%%%%%%%%%%%%%%%%%%%%%%%%%%%%%%%%%%%%%%%%%%%%%%%%%%%%%%%%%%%%%%%%%%
\section{Efficient Alternating Optimization Solution}
\label{sec:joint_opt}

We employ the alternating optimization approach. Specifically, we will alternately solve for $\ell$ and $\{\wv,\alphav\}$ while fixing the other, as presented in the following subsections.

\subsection{Optimization of $\ell$}
Given $\{\wv,\alphav\}$, $\ell$ is the solution to $\minimize\limits_{l \in \{0,1,\dots,L\}} ~\max(D^l,D^e)$. Notice that $D^l$ is monotonically decreases while $D^e$ monotonically increases with $\ell$. Therefore, $\max(D^l,D^e)$ is minimized when $D^l = D^e$, which holds for 
\begin{align} \label{eq:ell_opt}
	\ell = \hat{\ell}^\star = \frac{L \nu \RS(\wv, \alphav) f^e}{f^e f^l + \nu \RS(\wv, \alphav)(f^e + f^l)}.
\end{align}
We integerize $\ell$ and obtain the optimal offloading volume
\begin{align}
    \label{eq_sol_l}
	\ell^\star = \arg\min_{\hat{\ell} \in \{\lfloor \hat{\ell}^\star \rfloor, \lceil \hat{\ell}^\star \rceil\}} D(\wv, \alphav, \hat{\ell}).
\end{align}
From \eqref{eq:ell_opt}, we see that whenever the secrecy rate $\RS(\wv, \alphav)$ is zero, we have $\ell^\star = 0$. That is to say, if secure communication cannot be guaranteed, the user should execute the computation tasks locally.
%where $\hat{\ell}^*_k$ satisfies $D_k^l(\hat{\ell}^*_k) = D_k^e(\hat{\ell}^*_k)$, i.e., 

%\subsubsection{Optimization of $\fv^e$}
%Given $\Wm$, $\alphav$, and take $\ell_k = \hat{\ell}_k^*$, we optimize $\fv^e$ by
%\begin{align}
%	\min_{\fv^e}  &\sum_{k=1}^{K} \frac{\varpi_k (L_k c^2_k R_k + L_k c_k f_k^e)}{f_k^e f_k^l + c_k R_k(f_k^e + f_k^l)} \\
%	\text{subject to} ~~&\eqref{eq:constraint_ftot}, \eqref{eq:constraint_fe}.
%\end{align}
%This is a convex optimization problem and can be solved using the Lagrangian method as in \cite[Section III-A.2]{Bai2020latency}

\subsection{Joint Optimization of $\wv$ and $\alphav$}

We next jointly optimize $\wv$ and $\alphav$ given that $\ell = \ell^\star$. We have seen that with $\ell = \ell^\star$, it holds that $D^l \approx D^e$. Therefore, we can replace the objective function $D(\wv, \alphav, l)$ by $D^e$. In doing so, the optimization of $\wv$ and $\alphav$ is equivalent to the maximization of  the secrecy rate
\begin{align}
	(\Pc_{\rm secrecy}) ~~ \maximize_{\wv,\alphav} &~\RS(\wv_k,\alphav),
\end{align}
where it is implicit in this subsection that the constraints are \eqref{eq:constraint_alpha}, \eqref{eq:constraint_Pa} for fixed HRRIS and \eqref{eq:constraint_alpha}, \eqref{eq:constraint_Pa}, \eqref{eq:constraint_A} for dynamic HRRIS. 
Recall that $\RS(\wv,\alphav ) = (\REN(\wv,\alphav) - \barRE)^+$. The leakage rate $\barRE$ is independent of $(\wv,\alphav)$, and it is expected that for the optimal solution, $\REN(\wv,\alphav) > \barRE$. Therefore, we replace $\RS(\wv,\alphav )$ by $\REN(\wv_k,\alphav)$ in $(\Pc_{\rm secrecy})$, which leads to the maximization of the SINR, i.e.,
\begin{align*}
	(\Pc_{\rm SINR}) ~ \maximize_{\wv,\alphav} & \left(\!\gamma(\wv,\alphav) = \frac{P |\wv^\H(\hEN + \Gm \Upsilonm \hR)|^2}{\sigma^2 \wv^\H(\Id_M \!+\! \Gm\Psim\Psim^\H\Gm^\H) \wv}\!\right)\!.
\end{align*}

\subsubsection{Solution to $\wv$}
Given $\alphav$, the optimal $\wv$ maximizing the objective function in $(\Pc_{\rm SINR})$ can be found after some simple manipulations as
\begin{align}
    \label{eq_sol_w}
	\wv^\star = \sqrt{\frac{P}{\sigma^2}} (\Id_M + \Gm\Psim\Psim^\H\Gm^\H)^{-1} (\hEN + \Gm \Upsilonm \hR).
\end{align}

\subsubsection{Solution to $\alphav$}

In the case where there is no active element, i.e., $|\alpha_i| = 1, \forall i \in [N]$, the phases of $\alphav$ can be optimized as in~\cite[Sec.~IV]{Bai2020latency}. Specifically, the SINR becomes 
\begin{multline}
	\frac{P |\wv^\H(\hEN + \Gm \Upsilonm \hR)|^2}{\sigma^2 \|\wv\|^2} \\\le \frac{P}{\sigma^2 \|\wv\|^2} \bigg( |\wv^\H \hEN| + \sum_{n=1}^N |h_{{\rm H},n} \wv^\H \gv_n|   \bigg)^2, \label{eq:tmp304}
\end{multline}
where $h_{{\rm H},n}$ is the $n$-th entry of $\hR$ and $\gv_n$ is the $n$-th column of $\Gm$. The equality in \eqref{eq:tmp304} occurs if $\arg\{\wv^\H \hEN\} = \arg\{\wv^\H \Gm \Upsilonm \hR\}$. Accordingly, the phases of the HRRIS's coefficients can be obtained as
\begin{align}
	\boldsymbol{\theta}^{\star} = \arg\{\wv^\H \hEN\} - \arg\{\diag\{\wv^\H \Gm\} \hR\}, \label{eq:tmp308}
\end{align}
where $\arg\{\cdot\}$ returns the (element-wise) phases of a complex number or vector. When there is at least one active element, i.e., $A > 0$, both the phases and the amplitudes of $\{\alpha_i\}_{i\in \Ac}$ need to be optimized. We propose to first optimize the phases as in \eqref{eq:tmp308}, then optimize the amplitudes as follows.

\paragraph{Fixed HRRIS} 

The phases obtained in \eqref{eq:tmp308} result in $|\wv^\H(\hEN + \Gm \Upsilonm \hR)|^2 = \big( |\wv^\H \hEN| + \sum_{n=1}^N |\alpha_n h_{{\rm H},n} \wv^\H \gv_n|   \big)^2.$ Thus, the SINR can be expanded as
\begin{align}
	\frac{\gamma(\wv,\alphav)}{P/\sigma^2} %&= \frac{|\wv^\H \hEN|^2 + |\wv^\H \Gm \Upsilonm \hR|^2}{\wv^\H(\Id_M \!+\! \Gm\Psim\Psim^\H\Gm^\H) \wv} \\
	&= \frac{|\alpha_n|^2 a_n + |\alpha_n| b_n + c_n}{|\alpha_n|^2 u_n + v_n}, \label{eq:tmp311}
\end{align}
for any $n\in \Ac$, where
\begin{align}
	a_n &\defeq |h_{{\rm H},n}|^2 |\wv^\H \gv_n|^2, \\
	c_n &\defeq \bigg(|\wv^\H \hEN| + \sum_{i=1,i\ne n}^{N} |\alpha_i h_{{\rm H},i} \wv^\H \gv_i \bigg)^2, \\
	b_n &\defeq 2 |h_{{\rm H},n} \wv^\H \gv_n| \sqrt{c_n}, \\
	u_n &\defeq |\wv^\H \gv_n|^2, \\
	v_n &\defeq \|\wv\|^2 + \sum_{i\in \Ac, i\ne n} |\alpha_i|^2 |\wv^\H \gv_i|^2.
\end{align}
%with $h_{{\rm H},n}$ being the $n$-th entry of $\hR$ and $\gv_n$ the $n$-th column of $\Gm$. 
Constraint \eqref{eq:constraint_Pa} on the amplitudes is equivalent to
\begin{align} \label{eq:tmp324}
	P_{\rm a}(\alphav) = \sum_{n\in \Ac} |\alpha_n|^2 \xi_n \le P_{\rm a}^{\max}
\end{align}
where $\xi_n \defeq \sigma^2 + P |h_{{\rm H},n}|^2$. Let $\tilde{P}_{{\rm a},n} \defeq  \sum_{i\in \Ac, i\ne n} |\alpha_i|^2 \xi_i$, $n\in \Ac$, which is a constant if $\{|\alpha_i|\}_{i\in \Ac,n\ne n}$ are fixed. The constraint \eqref{eq:tmp324} can be written as $|\alpha_n| \le \sqrt{\frac{P_{\rm a}^{\max} - \tilde{P}_{{\rm a},n}}{\xi_n}}$. 
We alternatively optimize the amplitude of each active element while keeping the others fixed by solving
\begin{subequations} \label{eq_obj_fixedHRRIS}
	\begin{align}
			\maximize_{|\alpha_n|} &~  \frac{|\alpha_n|^2 a_n + |\alpha_n| b_n + c_n}{|\alpha_n|^2 u_n +  v_n}  \\ 
		\text{subject to} ~~&|\alpha_n| \le \sqrt{\frac{P_{\rm a}^{\max} - \tilde{P}_{{\rm a},n}}{\xi_n}}.
	\end{align}
\end{subequations}
The solution to \eqref{eq_obj_fixedHRRIS} is given in closed form as
\begin{align}
    \label{eq_sol_abs_alpha_fixedHRRIS}
    |\alpha_n|^\star = \min\left\{ \sqrt{\frac{P_{\rm a}^{\max} - \tilde{P}_{{\rm a},n}}{\xi_n}}, \frac{d_n}{b_n} + \sqrt{\frac{d_n^2}{b_n^2} + \frac{v_n}{u_n}}\right\},
\end{align}
where $d_n \defeq |h_{{\rm H},n}|^2 v_n - c_n$.
%This problem can be solved efficiently by analyzing the derivative of the objective function. 
We note that $|\alpha_n|^\star$ is not necessarily the maximal amplitude constraint $\sqrt{\frac{P_{\rm a}^{\max} - \tilde{P}_{{\rm a},n}}{\xi_n}}$.

\paragraph{Dynamic HRRIS}
For dynamic HRRIS, the set of active elements $\Ac$ subject to \eqref{eq:constraint_A} is also a design parameter. Determining both $\Ac$ and $\{|\alpha_i|\}_{i\in \Ac}$ is challenging since the active elements are involved in both the numerator and denominator of the SINR. Notice that 
\begin{align} 
	{\gamma}(\wv,\alphav)  \le \bar{\gamma}(\wv,\alphav) &\defeq \frac{P\big( |\wv^\H \hEN| \!+\! \sum_{n=1}^N\! |\alpha_n h_{{\rm H},n} \wv^\H \gv_n|   \big)^2}{\sigma^2\|\wv\|^2}.\! \notag \\
	&= \frac{P(|\alpha_n|^2 a_n + |\alpha_n| b_n + c_n)}{\sigma^2\|\wv\|^2} \label{eq:tmp340}
\end{align}
for any $n\in \Ac$.
The equality occurs when $\Ac = \emptyset$. The smaller the term $\sigma^2\wv^\H \Gm\Psim\Psim^\H\Gm^\H \wv$ is, the tighter the bound is. Observe that $\sigma^2\wv^\H \Gm\Psim\Psim^\H\Gm^\H \wv$ is small when the noise variance is small, the path loss is large, and/or the number of active elements is small. This motivates us to simplify the optimization to maximizing $\bar{\gamma}(\wv,\alphav)$. %, which can be written as $\frac{P(|\alpha_n|^2 a_n + |\alpha_n| b_n + c_n)}{\sigma^2\|\wv\|^2}$ for any $n\in \Ac$. 

An active element only improves the performance with respect to a passive one if $|\alpha_n| > 1$ since it causes performance loss due to signal attenuation otherwise. Thus, in the dynamic HRRIS, we have $|\alpha_n| > 1$, $\forall n \in \Ac$. Therefore, we ignore the $o(|\alpha_n|^2)$ terms, i.e., the terms involves $|\alpha_n|$ and the constant, in $\bar{\gamma}(\wv,\alphav)$ and focus on the second-order term. In doing so, we arrive at maximizing the terms $P a_n |\alpha_n|^2$. We consider the following problem 
\begin{subequations} \label{eq:opt_dyn}
	\begin{align}	
		\maximize_{\Ac,\{|\alpha_n|\}_{n\in \Ac}} &~\sum_{n\in \Ac} \log(1+P a_n |\alpha_n|^2), \\ 
		\text{subject to} ~~&\sum_{n\in \Ac} |\alpha_n|^2 \xi_n \le P_{\rm a}^{\max} \text{~and~} \eqref{eq:constraint_A}.
	\end{align}
\end{subequations}
Let $p_n \defeq |\alpha_n|^2 \xi_n$ denote the power allocated to the $n$-th active element, $n \in \Ac$. The optimization \eqref{eq:opt_dyn} can be rewritten as
\begin{subequations} \label{eq:opt_dyn_2}
	\begin{align}	
		\maximize_{\Ac,\{p_n\}_{n\in \Ac}} &~\sum_{n\in \Ac} \log\left(1+\frac{P a_n}{\xi_n} p_n\right), \\ 
		\text{subject to} ~~&\sum_{n\in \Ac} p_n \le P_{\rm a}^{\max} \text{~and~} \eqref{eq:constraint_A}.
	\end{align}
\end{subequations}
From \eqref{eq:opt_dyn_2}, we determine the optimal set of active elements $\Ac^\star$ as the set of indices of $A$ largest values in $\left\{ \frac{P a_1}{\xi_1},\dots,\frac{P a_N}{\xi_N} \right\}$. Given $\Ac^\star$, the optimal solution for $\{p_n\}$ is the well-known water-filling solution $p_n^\star = \left(\frac{1}{\mu} - \frac{\xi_n}{P a_n}\right)^+$ with $\mu$ satisfying  $\sum_{n\in \Ac} p_n \le P_{\rm a}^{\max}$. It follows that the amplitudes of the HRRIS's elements are given by
\begin{align}
    \label{eq_sol_abs_alpha}
	|\alpha^\star_n| = \begin{cases}
		\max \left\{\sqrt{\frac{1}{\mu \xi_n} - \frac{1}{P a_n}},1 \right\}, &\text{~for~} n \in \Ac^\star, \\
		1, &\text{~otherwise}.
	\end{cases}
\end{align}

\subsection{Overall Joint Computing and Communication Design}

The proposed joint computing and communication design scheme is summarized in Algorithm~\ref{alg_HRRIS_MEC}. We at first randomly generate the HRRIS coefficients satisfying constraints~\eqref{eq:constraint_alpha} and~\eqref{eq:constraint_Pa}. Then, in steps 2--11, the solutions to combining vector $\wv^{\star}$ and HRRIS coefficients $\alphav^{\star}$ are alternatively updated. More specifically, $\wv^{\star}$ is obtained based on~\eqref{eq_sol_w}, whereas, the HRRIS coefficients have the phases derived based on~\eqref{eq:tmp308} and the amplitudes derived in~\eqref{eq_sol_abs_alpha_fixedHRRIS} or~\eqref{eq_sol_abs_alpha} for the fixed or dynamic HRRIS schemes, respectively. Finally, when the update process terminated, i.e., when the convergence is reached, the optimal number of offloaded bits, i.e., $\ell^{\star}$, is obtained in step 12.

\begin{algorithm}[t]
	\small
	\caption{Latency minimization for HRRIS-aided secure MEC.}
	\label{alg_HRRIS_MEC}
	\begin{algorithmic}[1]
		\REQUIRE $\hE$, $\hEN$, $\hR$, and $\Gm$.
		\ENSURE $\{ \ell^{\star},\wv^{\star},\alphav^{\star} \}$.
		\STATE Randomly generate $\alphav$ satisfying constraints \eqref{eq:constraint_alpha} and \eqref{eq:constraint_Pa}.
		\WHILE{objective value has not converged}
    		\STATE Compute $\wv^{\star}$ based on \eqref{eq_sol_w}.
    		\STATE Compute $\boldsymbol{\theta}^{\star} = [\theta_1^{\star} \dots \theta_N^{\star}]$ based on \eqref{eq:tmp308}.
    		
    		\IF{fixed HRRIS is deployed}
    		    \STATE Obtain $\{|\alpha^\star_n|\}_{n\in \Ac}$ based on \eqref{eq_sol_abs_alpha_fixedHRRIS}. Set $|\alpha^\star_n| \!=\! 1$, $\forall n \!\in\! [N] \setminus \Ac$.  %for the fixed HRRIS.
    	    \ELSIF{dynamic HRRIS is deployed}
    	        \STATE Obtain $\{|\alpha^\star_n|\}_{n\in [N]}$ based on \eqref{eq_sol_abs_alpha}. % for the dynamic HRRIS.
            \ENDIF
    		\STATE Set $\alpha_n^{\star} = |\alpha^\star_n| e^{j \theta_n^{\star}}$, $n\in [N]$.
		\ENDWHILE
		\STATE Compute $\ell^{\star}$ based on \eqref{eq_sol_l}.
	\end{algorithmic}
\end{algorithm}

\section{Numerical Results}
\begin{figure}[t!]
	\vspace{-.5cm}
	\centering
	\includegraphics[width=.35\textwidth]{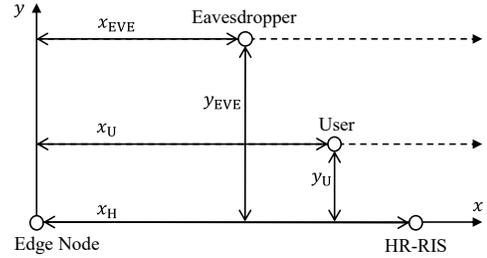}
	\vspace{-1cm}
	\caption{Locations of the EN, user, HRRIS, and eavesdropper.}
	\vspace{-0.3cm}
	\label{fig:scenario}
\end{figure}

In this section, we numerically evaluate the performance of the proposed HRRIS-assisted computation offloading schemes. We assume that uniform linear arrays (ULAs) are deployed at the EN, user, and eavesdropper, while a uniform planar array (UPA) of $N$ elements is deployed at the HRRIS. We assume half-wavelength distancing between array elements at all nodes. The locations of the nodes are illustrated in a two-dimensional coordinate in Fig.~\ref{fig:scenario}. The EN, HRRIS, user, and eavesdropper are located at $(0,0)$, $(x_{\rm H}, 0)$, $(x_{\rm H},y_{\rm H})$, and $(x_{\rm EVE},y_{\rm EVE})$, respectively. From these coordinates, the distance between the nodes can be easily computed using the Pythagorean theorem. 

For the large-scale fading, the path loss of a link distance $d$ is given by $\beta(d) = \beta_0\left(\frac{d}{1 {\rm m}}\right)^{-\eta}$, where $\beta_0$ is the path loss at the reference distance of $1$ meter (m), and $\eta$ is the path loss exponent. For small-scale fading, we assume the Rician fading channel model. Thus the small-scale fading channel between the HRRIS and the EN is modeled as $\tilde{\Gm} = \sqrt{\frac{\kappa}{1+\kappa}} \tilde{\Gm}^{\rm LoS} + \sqrt{\frac{1}{1+\kappa}} \tilde{\Gm}^{\rm NLoS}$, where $\tilde{\Gm}^{\rm LoS}$ and $\tilde{\Gm}^{\rm NLoS}$ represent the line-of-sight (LoS) and non-LoS~(NLoS) components, respectively, and $\kappa$ is the Rician factor. The NLoS component is modeled by the Rayleigh fading with independent entries following $\Cc\Nc(0,1)$. The LoS component is given by the product of the array response vectors which are computed from the azimuth and elevated angle-of-departure~(AoD) and angle-of-arrival~(AoA), as detailed in~\cite[Sec.~VI]{nguyen2021hybrid}.
%, i.e., $\tilde{\Gm}^{\rm LoS} = \av_{\rm EN}(\varphi_{EN})^\H \av_{\rm H} (\varphi_{\rm H},\phi_{\rm H})$. The $n$-th element of $\av_{\rm EN}(\varphi_{EN})$ and $\av_{\rm H} (\varphi_{\rm H},\phi_{\rm H})$ are given by $e^{\jmath \pi(n-1)\sin \varphi_{\rm EN}}$, $n \in [M]$, and $e^{\jmath \pi (\lfloor \frac{n}{N_x} \rfloor \sin \varphi_{\rm H} \sin \phi_{\rm H} + (n- \lfloor \frac{n}{N_x} \rfloor N_x) \sin \varphi_{\rm H} \cos \phi_{\rm H})}$, $n \in [N]$, respectively. Here $N_x$ is the number of elements in a horizontal dimension of the HRRIS. The angles $\varphi_{\rm H}, \varphi_{EN} \in [0,2\pi]$ denote the angle-of-departure~(AoD) at the HRRIS and the azimuth angle-of-arrival~(AoA) at the EN,
%respectively; $\phi\in [0,2\pi]$ denotes the elevation AoA
%at the HRRIS.
Finally, the channel between the HRRIS and the EN is obtained as $\Gm = \sqrt{\beta(x_{\rm H})} \tilde{\Gm}$. The channels of other links are modeled similarly.

For comparison, we consider the following schemes: 1) conventional passive RIS with random phase, 2) conventional passive RIS with optimized phase according to \eqref{eq:tmp308}, 3) optimized fixed HRRIS, and 4) optimized dynamic HRRIS. Note that the power consumption of the HRRIS schemes arises also from the active elements. For a fair comparison, we fix the same total power $P_{\rm tot} = P + P_{\rm a}^{\max}$ for all schemes. We also show the local computing latency $\frac{L\nu}{f^l}$ for reference. 

In the following, unless otherwise stated, we set $M = 5$, $E = 1$, $N = 50$, $A = 1$, $P_{\rm tot} = 30$~dBm, $P_{\rm a}^{\max} = 0$~dBm, $\sigma^2 = \sigma_{\rm EVE}^2 = -80$~dBm, $\{x_{\rm H}, x_{\rm U}, y_{\rm U}, x_{\rm EVE}, y_{\rm EVE}\} = \{50, 45, 2, 30, 9\}$~m, $\beta_0 = -30$ dB, $W = 1$~MHz, and $\epsilon = 0.1$. The path loss exponents of the user-EN, user-HRRIS, user-eavesdropper, and HRRIS-EN links are given by $3.5$, $2.2$, $2.8$, and $2.2$, respectively. The Rician factors of these links are set to $0, 1, 0$, and $100$, respectively. The computing parameters are given as $L = 300$~Kb, $\nu = 750$~CPU cycles/bit, $f^l = 5 \times 10^8$ cycles/s, and  $f^e = 20 \times 10^9$ cycles/s.% In each figure, a specified parameter is varied while the others remain fixed. All numerical results are averaged over $500$ independent channel realizations.

\begin{figure}[t!]
	\vspace{-.4cm}
	\centering
	\includegraphics[width=.48\textwidth]{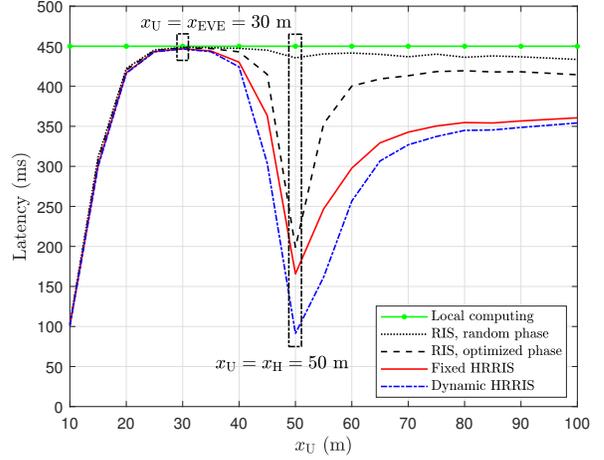}
	\vspace{-.4cm}
	\caption{Latency vs. location of the user.}
	\label{fig:latency_vs_xU}
	\vspace{-.2cm}
\end{figure}

First, in Fig.~\ref{fig:latency_vs_xU}, we investigate the latency achieved with the aforementioned schemes for different locations of the user. We vary $x_{\rm U}$ in $[10,100]$~m, i.e., let the user move along the $x$-axis in Fig.~\ref{fig:scenario}.  It is observed that the latency of all schemes is close to that of local computing when the user is closest to the eavesdropper. This confirms that when secure communication is not guaranteed, the user should only execute the computation tasks with local computing resources rather than exposing the data while offloading. On the other hand, when the user comes close to the EN or to the HRRIS, the latency induced by RIS with optimized phase or by HRRIS is minimized. This is because the direct user-EN link or the user-HRRIS link is strong, allowing the users to transmit at a high secrecy rate and offload most/all of the tasks to the computationally powerful EN. RIS with random phase only slightly reduces the latency even when the user is close to the RIS. The dynamic HRRIS outperforms the fixed HRRIS, and both significantly reduce the latency with respect to RIS, especially when the user is far from the EN. 

\begin{figure}[t!]
		\vspace{-.2cm}
	\centering
	\subfigure[Latency vs. $N$, $A=1$]{\includegraphics[width=.24\textwidth]{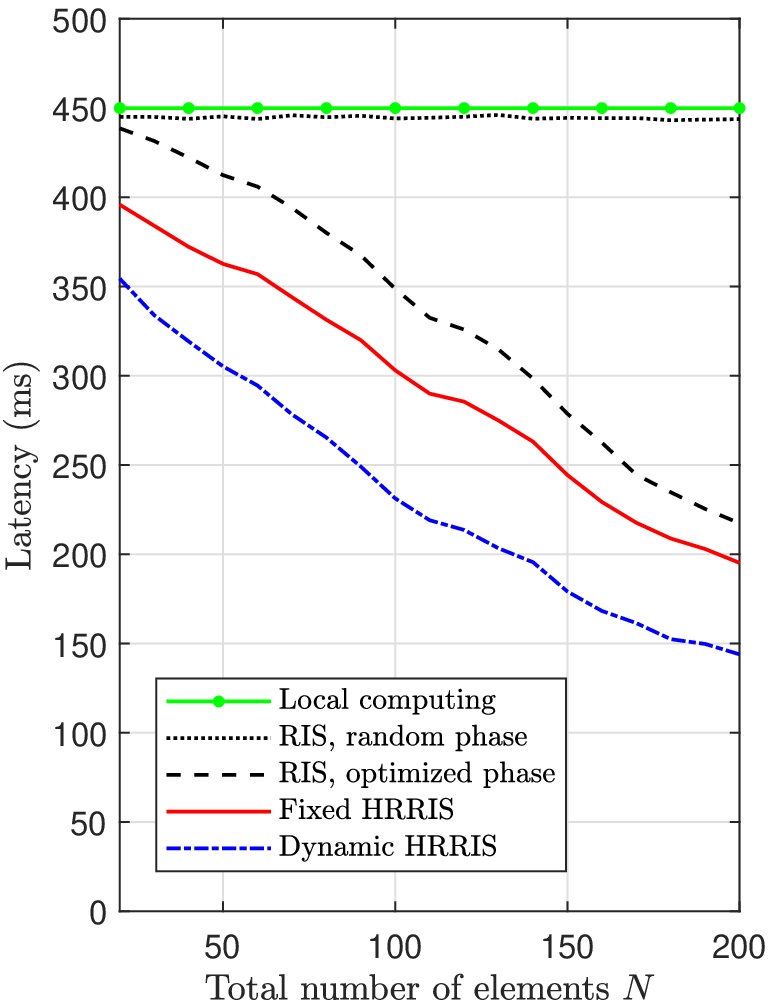}\label{fig:latency_vs_N}}
	\subfigure[Latency vs. $A$, $N =50$]{\includegraphics[width=.24\textwidth]{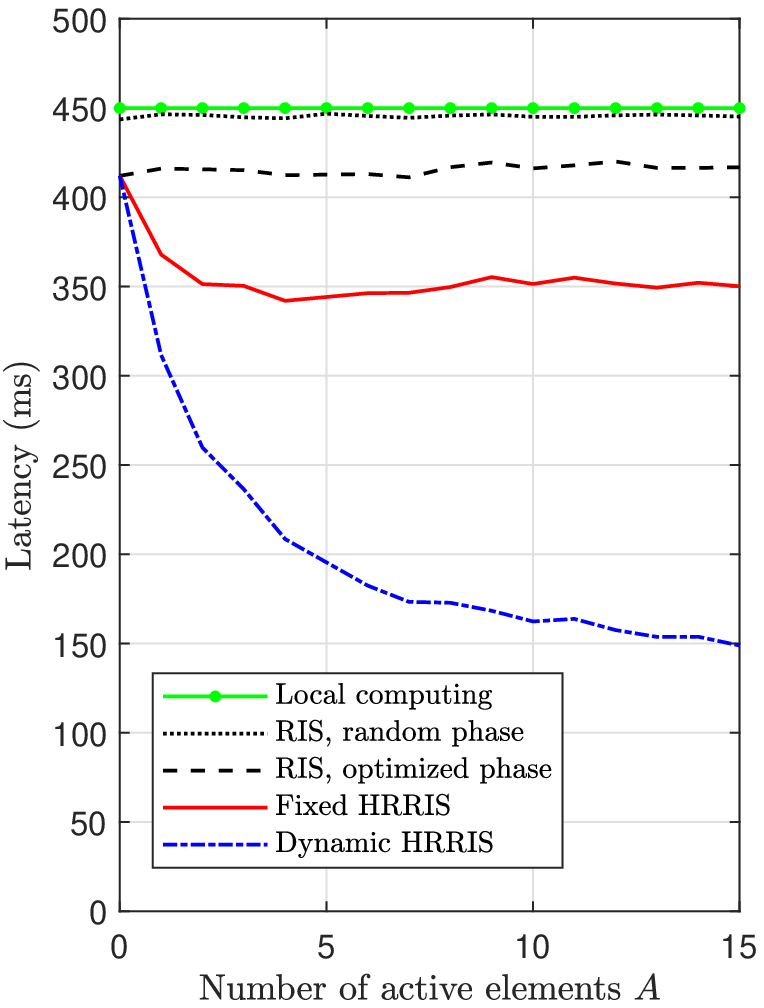}\label{fig:latency_vs_A}}
		\vspace{-.3cm}
	\caption{Latency vs. total number of RIS/HRRIS elements $N$ and the number of active elements $A$.}
	\vspace{-0.3cm}
	\label{fig:latency_vs_nElements}
\end{figure}

Next, in Fig.~\ref{fig:latency_vs_nElements}, we plot the latency as a function of $N$ and $A$. In Fig.~\ref{fig:latency_vs_N}, we observe again that RIS with random phase, even with a large number of elements, only slightly improves upon local computing. The latency induced by RIS with optimized phase and by HRRIS decreases as $N$ increases. Interestingly, while the gain of the fixed HRRIS with respect to RIS with optimized phase shrinks for large $N$, the dynamic HRRIS can maintain a high gain. Fig.~\ref{fig:latency_vs_A} shows that the latency of both the fixed and dynamic HRRIS saturate as the number of active elements $A$ increases, but the floor for the dynamic HRRIS is much lower than that for the fixed HRRIS.

\begin{figure}[t!]
		\vspace{-.2cm}
	\centering
	\subfigure[Latency vs. $f^e$, $f^l = 5 \times 10^8$]{\includegraphics[width=.24\textwidth]{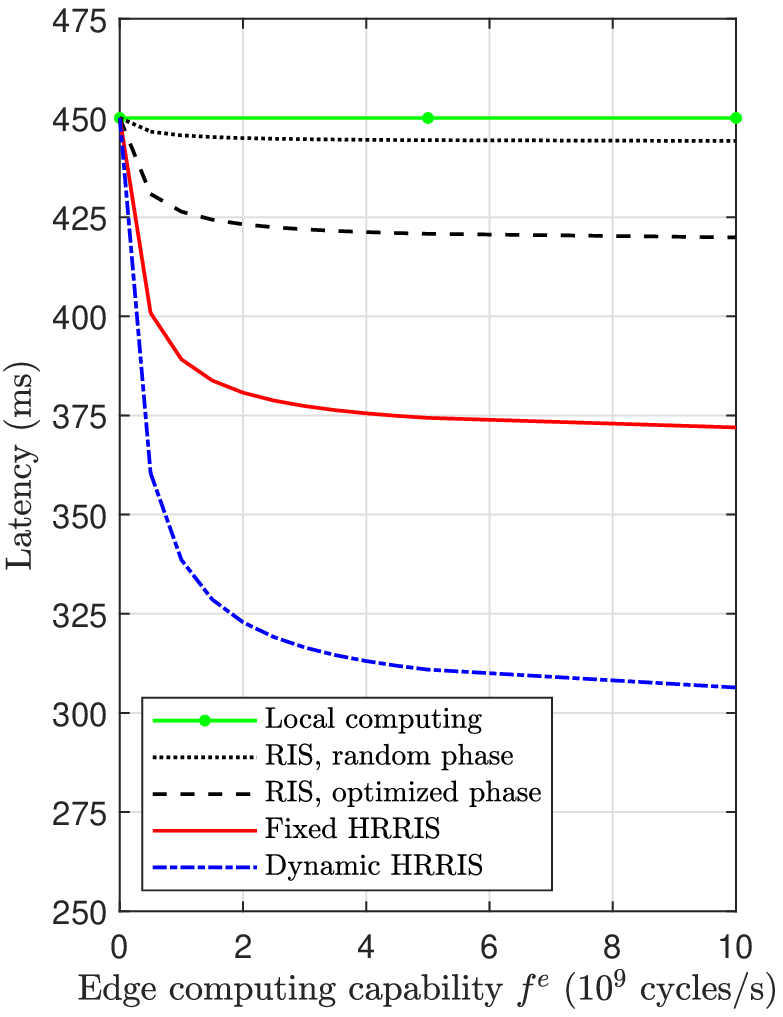}\label{fig:latency_vs_fe}}
	\subfigure[Latency vs. $f^l$, $f^e = 20 \times 10^9$]{\includegraphics[width=.24\textwidth]{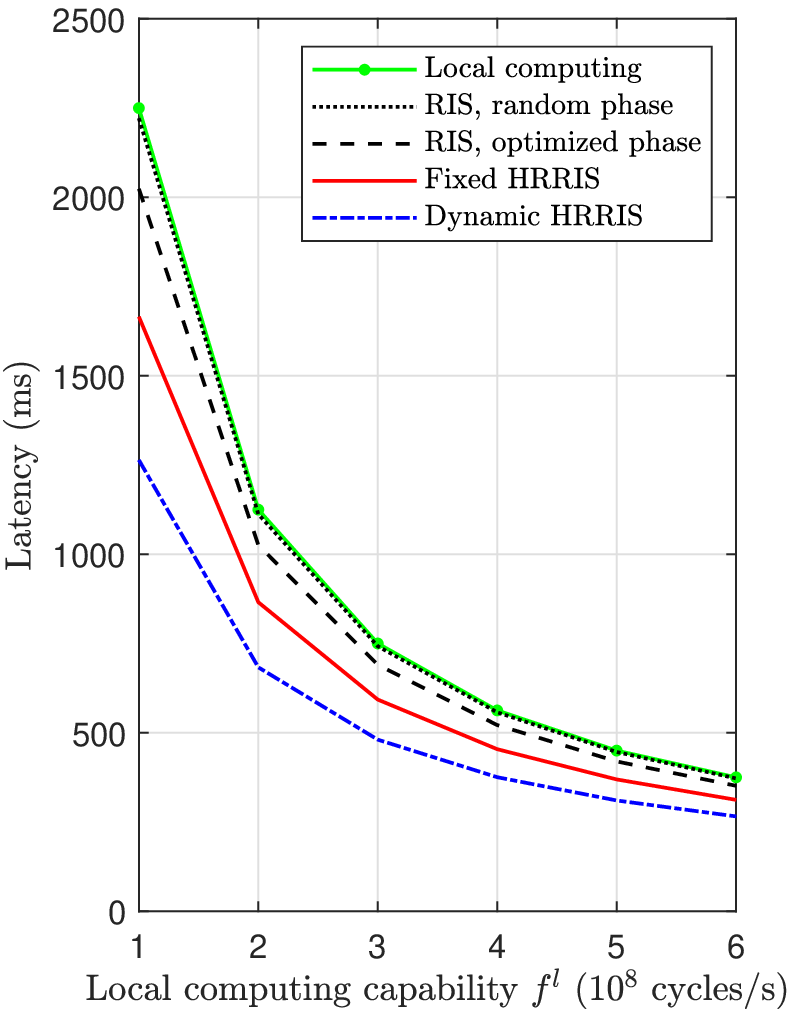}\label{fig:latency_vs_fl}}
		\vspace{-.3cm}
	\caption{Latency vs. edge computing capability $f^e$ and local computing capability $f^l$. The EN has $M=5$ antennas.}
	\vspace{-0.3cm}
	\label{fig:latency_vs_comp}
\end{figure}

Furthermore, we investigate the latency improvement of the HRRIS when the edge/local computing capability varies in Fig.~\ref{fig:latency_vs_comp}. We still observe that the HRRIS achieves the lowest latency. In Fig.~\ref{fig:latency_vs_fe}, the latency is reduced drastically when the edge computing capability $f^e$ increases from a small value, whereas only a minor reduction in the latency can be seen when $f^e$ is large. This has been observed for the RIS in~\cite{Bai2020latency}. The reason is that the latency imposed by edge computing dominates when $f^e$ is small, whereas the latency imposed by communication dominates when  $f^e$ is large. In Fig.~\ref{fig:latency_vs_fl}, we see that the latency of all schemes decreases proportional to $\frac{1}{f^l}$. The relative gain of HRRIS remains as $f^l$ increases.

\begin{figure}[t!]
	\vspace{-.1cm}
	\centering
	\subfigure[$P_{\rm a}^{\rm max} = 0$~dBm]{\includegraphics[width=.24\textwidth]{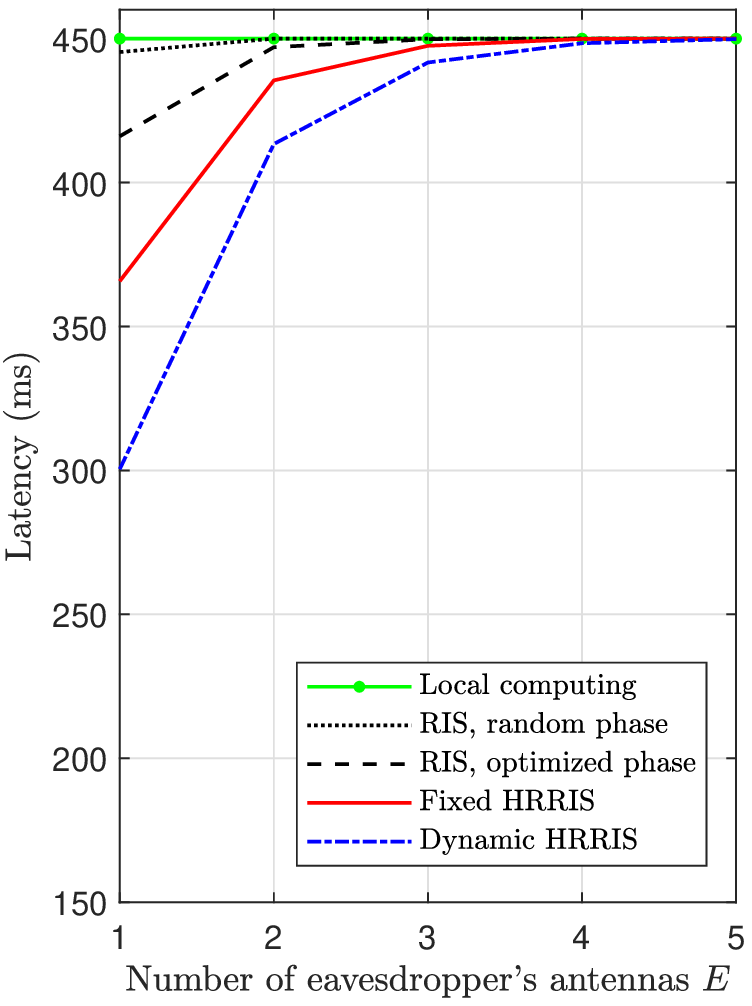}}
	\subfigure[$P_{\rm a}^{\rm max} = 10$~dBm]{\includegraphics[width=.24\textwidth]{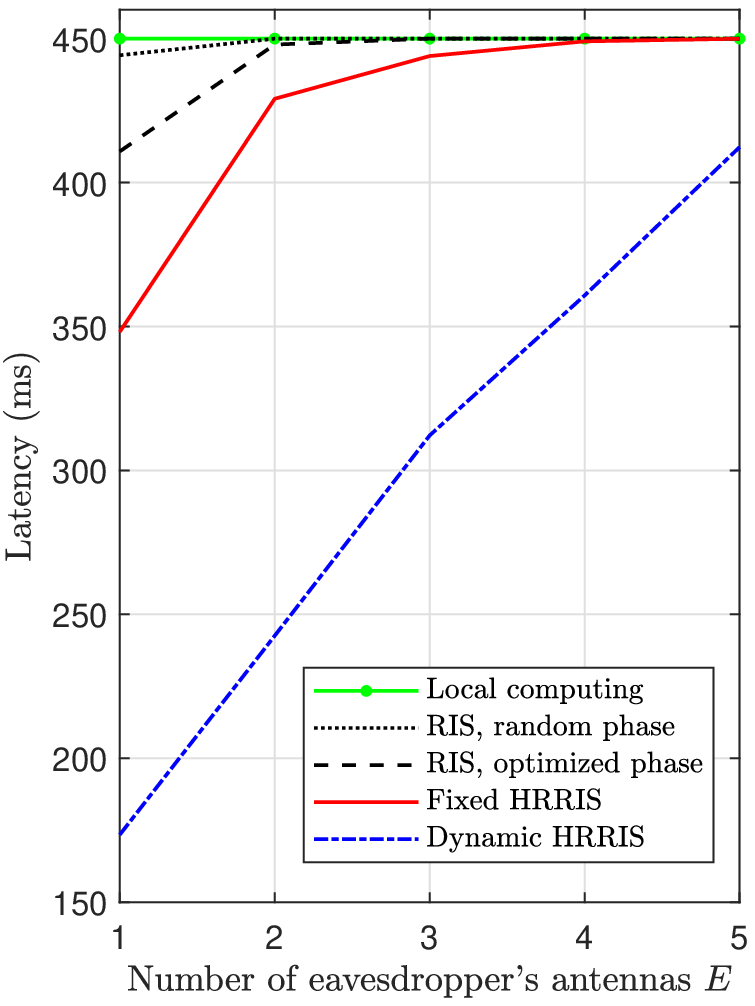}}
	\vspace{-.3cm}
	\caption{Latency vs. number of eavesdropper's antennas $E$.}
	\vspace{-0.4cm}
	\label{fig:latency_vs_E}
\end{figure}

Finally, Fig.~\ref{fig:latency_vs_E} depicts the latency as a function of the number of eavesdropper's antennas $E$ for two values of $P_{\rm a}^{\max}$, namely, $0$ and $10$~dBm. Naturally, the latency of all schemes increases as the eavesdropper becomes more capable. When $P_{\rm a}^{\max} = 0$~dBm, none of the schemes can improve the latency upon local computing if the eavesdropper has a comparable number of antennas as the EN. The situation remains similar for RIS and fixed HRRIS if $P_{\rm a}^{\max}$ goes up to $10$~dBm. However, the dynamic HRRIS can effectively exploit the extra power to reduce significantly the latency, achieving approximately $9$~\% gain with respect to other schemes for $P_{\rm a}^{\max} = 10$~dBm when the eavesdropper has as many antennas as the EN.

\section{Conclusion}
In this paper, we investigate  a single-user MEC system assisted by an HRRIS. We aim at minimizing the user's offloading latency while ensuring secure transmission against a malicious eavesdropper. % investigate the use of intelligence surface to assist low-latency and secure computation offloading in a single-user mobile edge computing system with an eavesdropper.
To this end, we develop a novel alternating optimization method by sequentially optimizing the surface's configuration, the receive combining vector, and the computation offloading volume. By exploring the structure of HRRIS, our numerical results show that introducing as few as one active relaying element in the surface can reduce significantly the latency in comparison with conventional RIS, especially if the position of the active element can be dynamically changed. Possible future extensions include investigating a multi-user system where resource allocation between the users also plays a key role. 

\bibliographystyle{IEEEtran}
\bibliography{IEEEAbbriv,./biblio}

\begin{thebibliography}{10}
\providecommand{\url}[1]{#1}
\csname url@samestyle\endcsname
\providecommand{\newblock}{\relax}
\providecommand{\bibinfo}[2]{#2}
\providecommand{\BIBentrySTDinterwordspacing}{\spaceskip=0pt\relax}
\providecommand{\BIBentryALTinterwordstretchfactor}{4}
\providecommand{\BIBentryALTinterwordspacing}{\spaceskip=\fontdimen2\font plus
\BIBentryALTinterwordstretchfactor\fontdimen3\font minus
  \fontdimen4\font\relax}
\providecommand{\BIBforeignlanguage}[2]{{%
\expandafter\ifx\csname l@#1\endcsname\relax
\typeout{** WARNING: IEEEtran.bst: No hyphenation pattern has been}%
\typeout{** loaded for the language `#1'. Using the pattern for}%
\typeout{** the default language instead.}%
\else
\language=\csname l@#1\endcsname
\fi
#2}}
\providecommand{\BIBdecl}{\relax}
\BIBdecl

\bibitem{LA20193}
Q.~D. La, M.~V. Ngo, T.~Q. Dinh, T.~Q. Quek, and H.~Shin, ``Enabling
  intelligence in fog computing to achieve energy and latency reduction,''
  \emph{Digital Communications and Networks}, vol.~5, no.~1, pp. 3--9, Sep.
  2019, {A}rtificial Intelligence for Future Wireless Communications and
  Networking.

\bibitem{Jiang2019toward}
C.~Jiang, X.~Cheng, H.~Gao, X.~Zhou, and J.~Wan, ``Toward computation
  offloading in edge computing: {A} survey,'' \emph{IEEE Access}, vol.~7, pp.
  131\,543--131\,558, Aug. 2019.

\bibitem{Bai2020latency}
T.~Bai, C.~Pan, Y.~Deng, M.~Elkashlan, A.~Nallanathan, and L.~Hanzo, ``Latency
  minimization for intelligent reflecting surface aided mobile edge
  computing,'' \emph{IEEE J. Sel. Topics Signal Process.}, vol.~38, no.~11, pp.
  2666--2682, Nov. 2020.

\bibitem{di2020smart}
{M. Di Renzo \textit{et al.}}, ``Smart radio environments empowered by
  reconfigurable intelligent surfaces: {H}ow it works, state of research, and
  the road ahead,'' \emph{{IEEE} J. Sel. Areas Commun.}, vol.~38, no.~11, pp.
  2450--2525, Nov. 2020.

\bibitem{he2021channel}
J.~He, N.~T. Nguyen, R.~Schroeder, V.~Tapio, J.~Kokkoniemi, and M.~Juntti,
  ``Channel estimation and hybrid architectures for {RIS}-assisted
  communications,'' \emph{arXiv preprint arXiv:2104.07115}, 2021.

\bibitem{wu2020intelligent}
Q.~Wu, S.~Zhang, B.~Zheng, C.~You, and R.~Zhang, ``{Intelligent reflecting
  surface aided wireless communications: A tutorial},'' \emph{{IEEE} Trans.
  Commun.}, vol.~69, no.~5, pp. 3313--3351, Jan. 2021.

\bibitem{nguyen2021machine}
N.~T. Nguyen, L.~V. Nguyen, T.~Huynh-The, D.~H. Nguyen, A.~L. Swindlehurst, and
  M.~Juntti, ``Machine learning-based reconfigurable intelligent surface-aided
  {MIMO} systems,'' \emph{arXiv preprint arXiv:2105.00347}, 2021.

\bibitem{liu2021reconfigurable}
Y.~Liu, X.~Liu, X.~Mu, T.~Hou, J.~Xu, M.~Di~Renzo, and N.~Al-Dhahir,
  ``{Reconfigurable intelligent surfaces: Principles and opportunities},''
  \emph{{IEEE} Commun. Surveys Tuts.}, May 2021, early access.

\bibitem{nguyen2021hybrid_mag}
{N. T. Nguyen \textit{et al.}}, ``Hybrid relay-reflecting intelligent
  surface-aided wireless communications: Opportunities, challenges, and future
  perspectives,'' \emph{arXiv preprint arXiv:2104.02039}, 2021.

\bibitem{nguyen2021spectral}
N.~T. Nguyen, Q.-D. Vu, K.~Lee, and M.~Juntti, ``Spectral efficiency
  optimization for hybrid relay-reflecting intelligent surface,'' in
  \emph{Proc. IEEE Int. Conf. Commun. Workshop}, 2021.

\bibitem{nguyen2021hybrid}
------, ``{Hybrid Relay-Reflecting Intelligent Surface-Assisted Wireless
  Communication},'' \emph{arXiv preprint arXiv:2103.03900}, 2021.

\bibitem{Hu2021reconfigurable}
X.~Hu, C.~Masouros, and K.-K. Wong, ``Reconfigurable intelligent surface aided
  mobile edge computing: From optimization-based to location-only
  learning-based solutions,'' \emph{{IEEE} Trans. Commun.}, Mar. 2021.

\bibitem{khisti2010secure}
A.~Khisti and G.~W. Wornell, ``{Secure transmission with multiple
  antennas—Part II: The MIMOME wiretap channel},'' \emph{IEEE Trans. Inf.
  Theory}, vol.~56, no.~11, pp. 5515--5532, Nov. 2010.

\bibitem{liu2014secrecy}
L.~Liu, R.~Zhang, and K.-C. Chua, ``{Secrecy wireless information and power
  transfer with MISO beamforming},'' \emph{IEEE Trans. Signal Process.},
  vol.~62, no.~7, pp. 1850--1863, Apr. 2014.

\bibitem{mukherjee2014principles}
A.~Mukherjee, S.~A.~A. Fakoorian, J.~Huang, and A.~L. Swindlehurst,
  ``{Principles of physical layer security in multiuser wireless networks: A
  survey},'' \emph{{IEEE} Commun. Surveys Tuts.}, vol.~16, no.~3, pp.
  1550--1573, Feb. 2014.

\bibitem{cui2019secure}
M.~Cui, G.~Zhang, and R.~Zhang, ``Secure wireless communication via intelligent
  reflecting surface,'' \emph{{IEEE} Wireless Commun. Lett.}, vol.~8, no.~5,
  pp. 1410--1414, Oct. 2019.

\bibitem{chen2019intelligent}
J.~Chen, Y.-C. Liang, Y.~Pei, and H.~Guo, ``{Intelligent reflecting surface: A
  programmable wireless environment for physical layer security},'' \emph{IEEE
  Access}, vol.~7, pp. 82\,599--82\,612, Jun. 2019.

\bibitem{hong2020artificial}
S.~Hong, C.~Pan, H.~Ren, K.~Wang, and A.~Nallanathan, ``Artificial-noise-aided
  secure {MIMO} wireless communications via intelligent reflecting surface,''
  \emph{{IEEE} Trans. Commun.}, vol.~68, no.~12, pp. 7851--7866, Dec. 2020.

\bibitem{yu2020robust}
X.~Yu, D.~Xu, Y.~Sun, D.~W.~K. Ng, and R.~Schober, ``Robust and secure wireless
  communications via intelligent reflecting surfaces,'' \emph{{IEEE} J. Sel.
  Areas Commun.}, vol.~38, no.~11, pp. 2637--2652, Nov. 2020.

\bibitem{shen2019secrecy}
H.~Shen, W.~Xu, S.~Gong, Z.~He, and C.~Zhao, ``Secrecy rate maximization for
  intelligent reflecting surface assisted multi-antenna communications,''
  \emph{{IEEE} Commun. Lett.}, vol.~23, no.~9, pp. 1488--1492, Sep. 2019.

\bibitem{wu2018intelligent}
Q.~Wu and R.~Zhang, ``{Intelligent reflecting surface enhanced wireless
  network: Joint active and passive beamforming design},'' in \emph{Proc. IEEE
  Global Commun. Conf.}, Dec. 2018, pp. 1--6.

\bibitem{Khandaker2018}
M.~R.~A. Khandaker, C.~Masouros, and K.-K. Wong, ``Constructive interference
  based secure precoding: {A} new dimension in physical layer security,''
  \emph{{IEEE} Trans. Inf. Forensics Security}, vol.~13, no.~9, pp. 2256--2268,
  Sep. 2018.

\bibitem{He2020}
X.~He, R.~Jin, and H.~Dai, ``Physical-layer assisted secure offloading in
  mobile-edge computing,'' \emph{{IEEE} Trans. Wireless Commun.}, vol.~19,
  no.~6, pp. 4054--4066, Jun. 2020.

\end{thebibliography}

\end{document}